\documentclass{kapproc} 
\usepackage{procps} 
\usepackage[dvips,dvipdf]{graphicx}
\graphicspath{{graphicss}}
\upperandlowercase
\kluwerbib 

\begin{document}

\articletitle{Photometric and spectroscopic follow-up of optical counterparts of X-ray 
 sources in the RasTyc sample}

\author{Antonio Frasca,\altaffilmark{1} Patrick Guillout,\altaffilmark{2} 
Rubens Freire Ferrero,\altaffilmark{2}  Ettore Marilli,\altaffilmark{1} and
Katia Biazzo\altaffilmark{1}}

\altaffiltext{1}{INAF - Osservatorio Astrofisico di Catania\\ 
via S.~Sofia 78\\ I--95123 Catania, Italy}
\email{afrasca@ct.astro.it}

\altaffiltext{2}{Observatoire Astronomique\\ 11 rue de l'Universite\\
67000 Strasbourg, France}
\email{guillout@newb6.u-strasbg.fr}

\chaptitlerunninghead{Optical counterparts of X-ray 
 sources in the RasTyc sample}

\begin{abstract}
The cross-correlation between the ROSAT all-sky survey 
($\sim 150\,000$ sources) and the Tycho mission 
($\sim 1\,000\,000$ stars) catalogs has selected about 14\,000 
stellar X-ray sources (RasTyc sample, Guillout et al. 1999).
 
About 200-300 stars have been spectroscopically observed 
at high resolution both in the  H$\alpha$ and Li{\sc i} $\lambda$6708
regions with E{\sc lodie} and A{\sc urelie} spectrographs of the OHP
(Observatoire de l'Haute Provence, France). The aim was to classify the
RasTyc star sample in terms of age and chromospheric activity level
and to detect eventual binary systems.
 
Photometric and spectroscopic follow-up observations of a RasTyc sub-sample
composed of particularly interesting objects (binaries and very young stars)
has been performed with the 91-cm telescope of the Catania Astrophysical
Observatory.
 
In this work we present some results of this monitoring.
In particular, we have obtained good radial velocity curves and solved for 
the orbits of three SB2 and three SB1 spectroscopic binaries. 
In addition, for near all sources we have detected a photometric 
modulation ascribable to photospheric surface inhomogeneities and 
chromospheric H$\alpha$ line variation. \\
\end{abstract}


\section*{Introduction}

From the ROSAT All-Sky Survey ($\simeq$ 150\,000 sources) and the TYCHO catalogs
($\simeq$ 1\,000\,000 stars) Guillout et al. (1999) selected about 14\,000 optical counterparts of
stellar X-ray sources, the so-called RasTyc sample. 
The high X-ray emission level of these sources is compatible with young stars
in the solar neighborhoods, or with close binaries of the RS~CVn and BY~Dra classes, for which
the high magnetic activity level is due to the synchronization of rotational and orbital periods.
The knowledge of the incidence of binaries in this sample is of paramount importance for studying the recent 
local star formation history. We thus have started a spectroscopic observation campaign 
aimed at a deep characterization of a representative sub-sample of the RasTyc population. 
As a matter of fact, the high-resolution spectroscopic observations allow to derive spectral types and 
luminosity classes, to infer ages (by means of Lithium abundance) and to single out spectroscopic binaries. 
In addition, chromospheric activity levels (from H$\alpha$ emission) and rotational 
velocities (from Doppler broadening), can be also determined. 

In this work we report on the discovery of three single-lined (SB1) and three double-lined (SB2) late-type
binaries, which have been observed with the 193-cm telescope of the {\it Observatoire de Haute Provence} (OHP) 
and the 91-cm telescope of the Catania Astrophysical Observatory (OAC). 
For these binaries we have obtained good radial velocity 
curves and orbital solutions. An accurate spectral classification has been 
also performed and the projected rotational velocity $v\sin i$ has been measured for all stars. 
The  H$\alpha$ emission  and the Li{\sc i}\,$\lambda\,6708$ line have been used to infer the chromospheric 
activity level and the lithium content, respectively.

\section{Observations and reduction}
\label{sec:Obs}

\subsection{Spectroscopy}

Spectroscopic observations have been obtained at the {\it Observatoire de Haute Provence} 
(OHP) and at the {\it M.G. Fracastoro} station (Serra La Nave, Mt. Etna, 1750 m a.s.l.) of the Catania Astrophysical 
Observatory (OAC).

At OHP we observed in 2000 and 2001 with the E{\sc lodie} echelle spectrograph connected to the 193-cm 
telescope.  The 67 orders recorded by the CCD detector cover the 3906-6818~\AA\ 
wavelength range with a resolving power of about 42\,000 (Baranne et al. 1996). The E{\sc lodie} 
spectra were automatically reduced on-line during the observations and the cross-correlation 
with a reference mask was produced as well.
 
The observations carried out at Catania Observatory have been performed in 2001
and 2002 with the REOSC echelle spectrograph at the 91-cm telescope. 
The spectrograph is fed by the telescope through an optical fiber
(UV - NIR, $200\,\mu m$ core diameter) and is placed in a stable position
in the room below the dome level. Spectra were recorded on a CCD camera 
equipped with a thinned back-illuminated
SITe CCD of 1024$\times$1024 pixels (size 24$\times$24 $\mu$m). The echelle crossed 
configuration yields a resolution of about 14\,000, as deduced from the FWHM 
of the lines of the Th-Ar calibration lamp. The observations have been made in the red 
region. The detector allows us to record five orders in each frame, spanning from about 5860
to 6700~\AA. 

The OAC data reduction was performed by using the {\sc echelle} task of IRAF\footnote{IRAF is 
distributed by the National Optical Astronomy Observatories,
which are operated by the Association of Universities for Research
in Astronomy, Inc., under cooperative agreement with the National
Science Foundation.} 
package following the standard steps: background subtraction, division by a flat
field spectrum given by a halogen lamp, wavelength calibration using the 
emission lines of a Th-Ar lamp, and normalization to the continuum through a polynomial fit.

\subsection{Photometry}
The photometric observations have been carried out in 2001 and 2002 in the standard $UBV$ 
system with the 91-cm telescope of OAC and a photon-counting refrigerated photometer 
equipped with an EMI 9789QA photomultiplier, cooled to $-15^\circ$C. The dark noise of the detector, 
operated at this temperature, is about $1$ photon/sec.

For each field of the RasTyc sources, we have chosen two or three stars with known $UVB$ magnitudes 
to be used as local standards for the determination of the photometric instrumental ``zero points''. 
Additionally, several standard stars, selected from the list of Landolt (1992), 
were also observed during the run in order to determine the transformation coefficients 
to the Johnson standard system. 

A typical observation consisted of several integration cycles (from 1 to 3, depending on the star brightness) 
of 10, 5, 5 seconds, in the $U$, $B$ and $V$ filter, respectively.
A 21-arcsecond diaphragm was used. The data were reduced by means of the photometric data
reduction package PHOT designed for photoelectric photometry of Catania Observatory 
(Lo Presti \& Marilli 1993). Seasonal mean extinction coefficient for Serra La Nave Observatory were adopted for 
the atmospheric extinction correction.

\section{Results}

\subsection{Radial velocity and photometry}
\label{sec:RV}

\begin{figure*}[ht]
\begin{center}
\includegraphics[width=5.9cm,height=4.6cm]{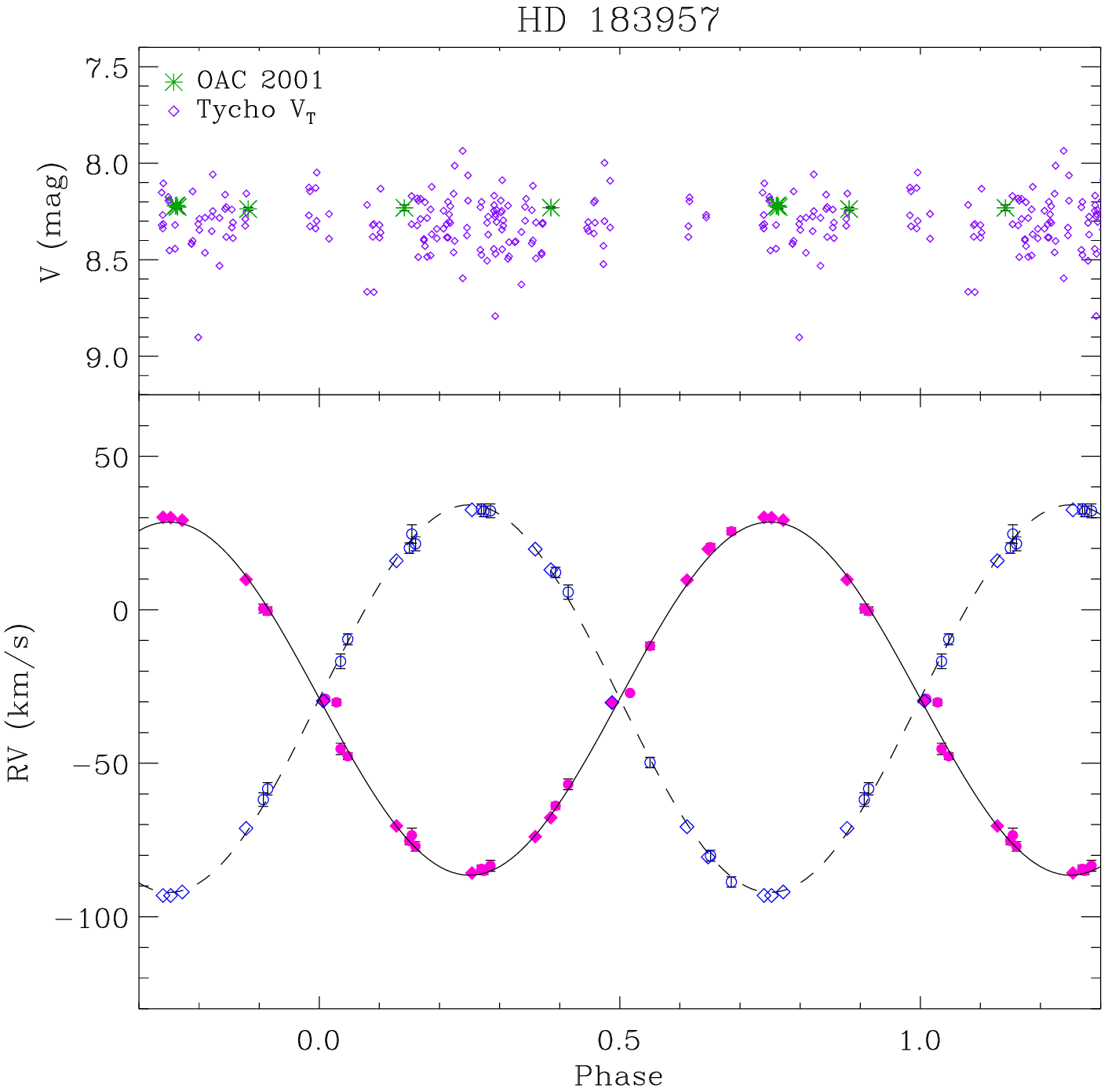}  
\includegraphics[width=5.9cm,height=4.6cm]{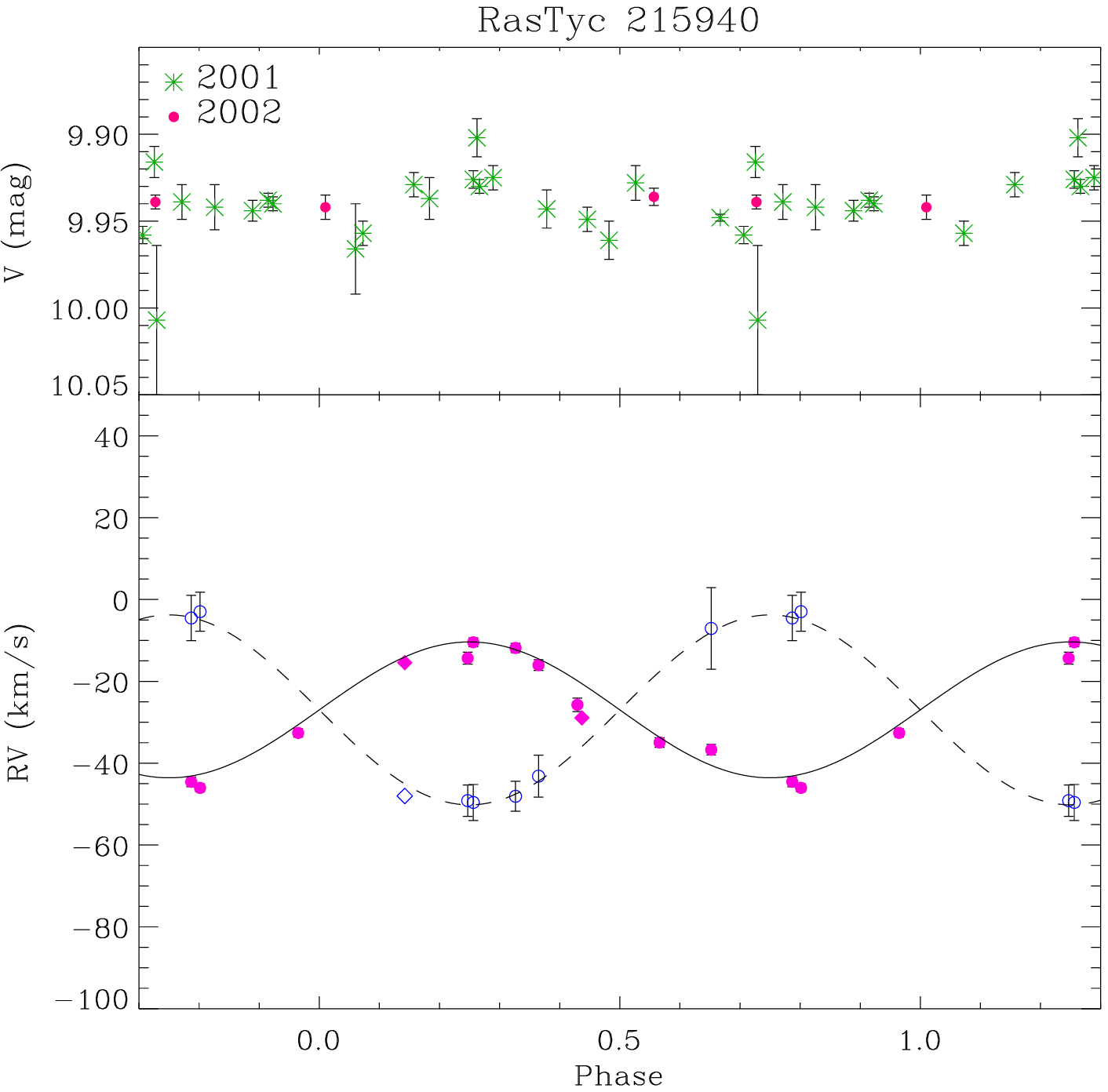}  
\includegraphics[width=5.9cm,height=4.6cm]{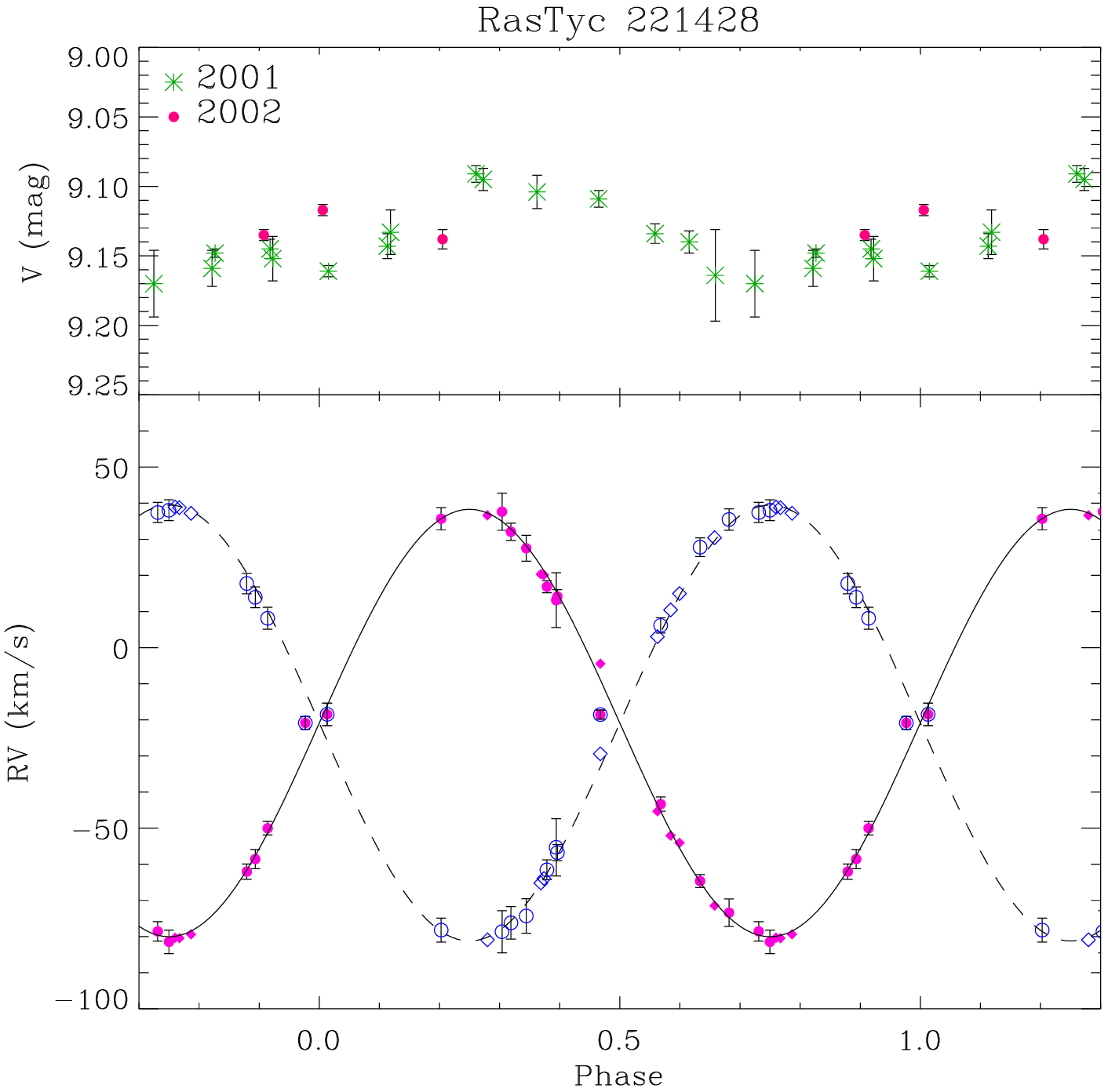}  
\includegraphics[width=5.9cm,height=4.6cm]{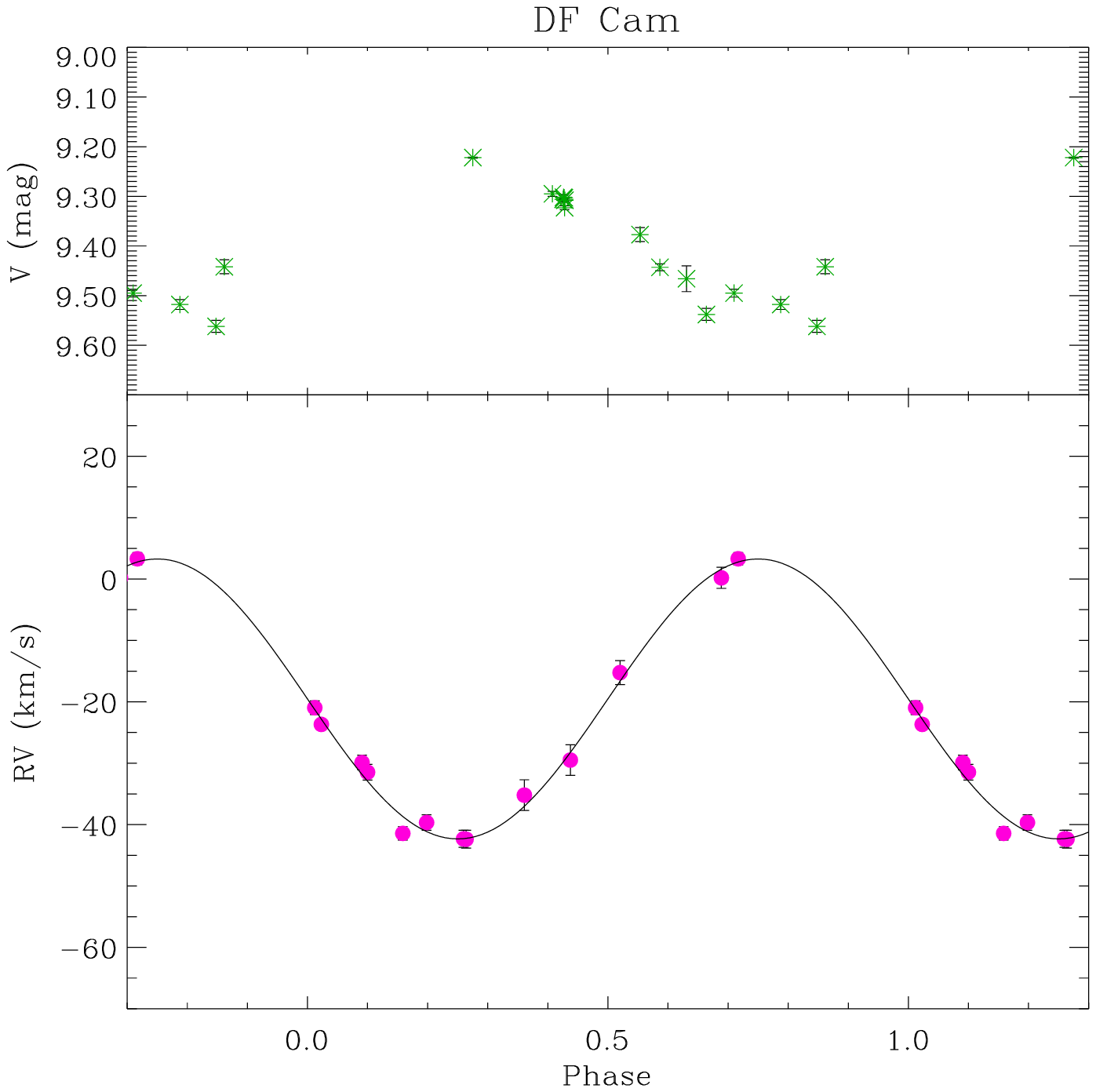}  
\includegraphics[width=5.9cm,height=4.6cm]{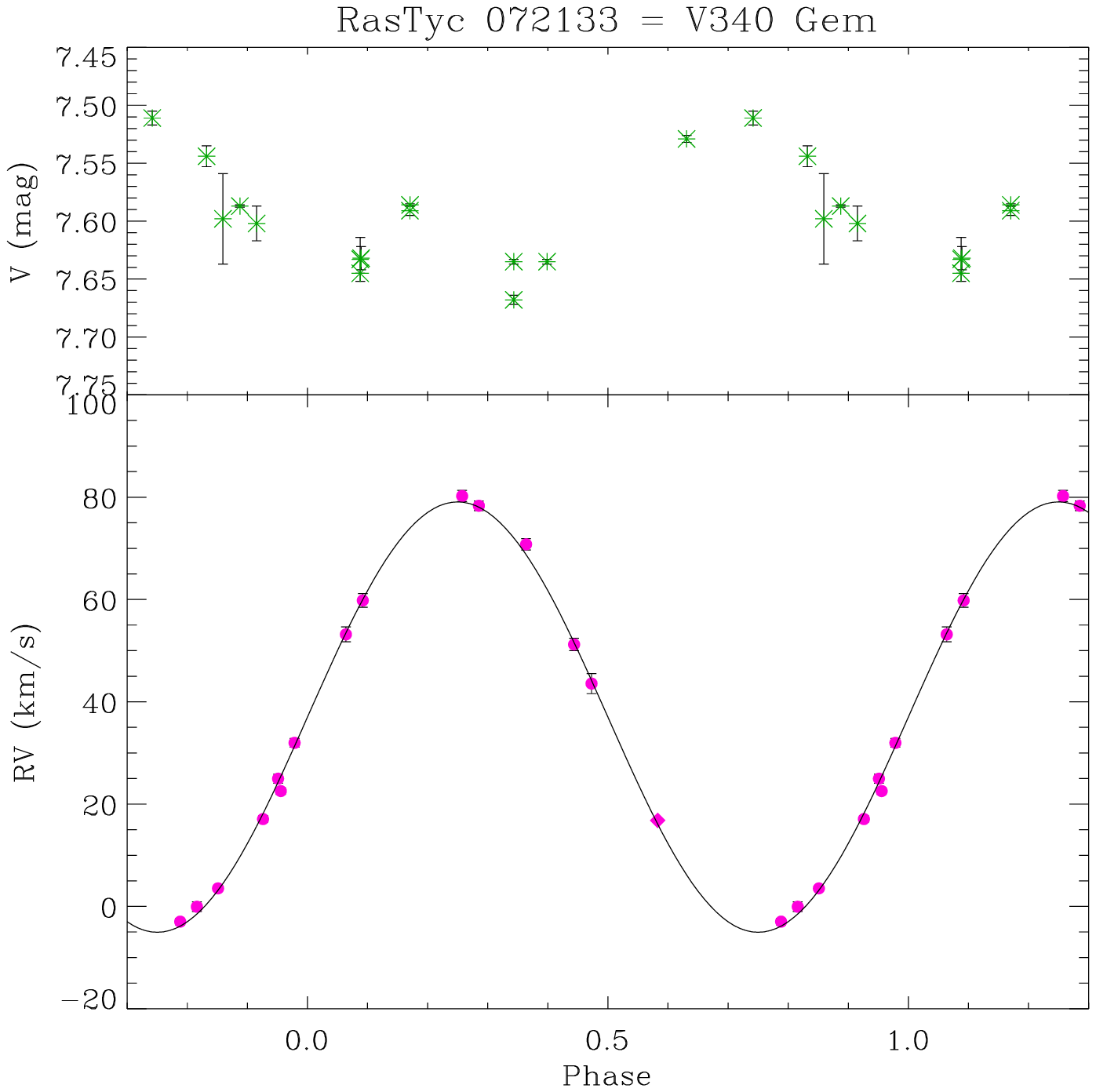}  
\includegraphics[width=5.9cm,height=4.6cm]{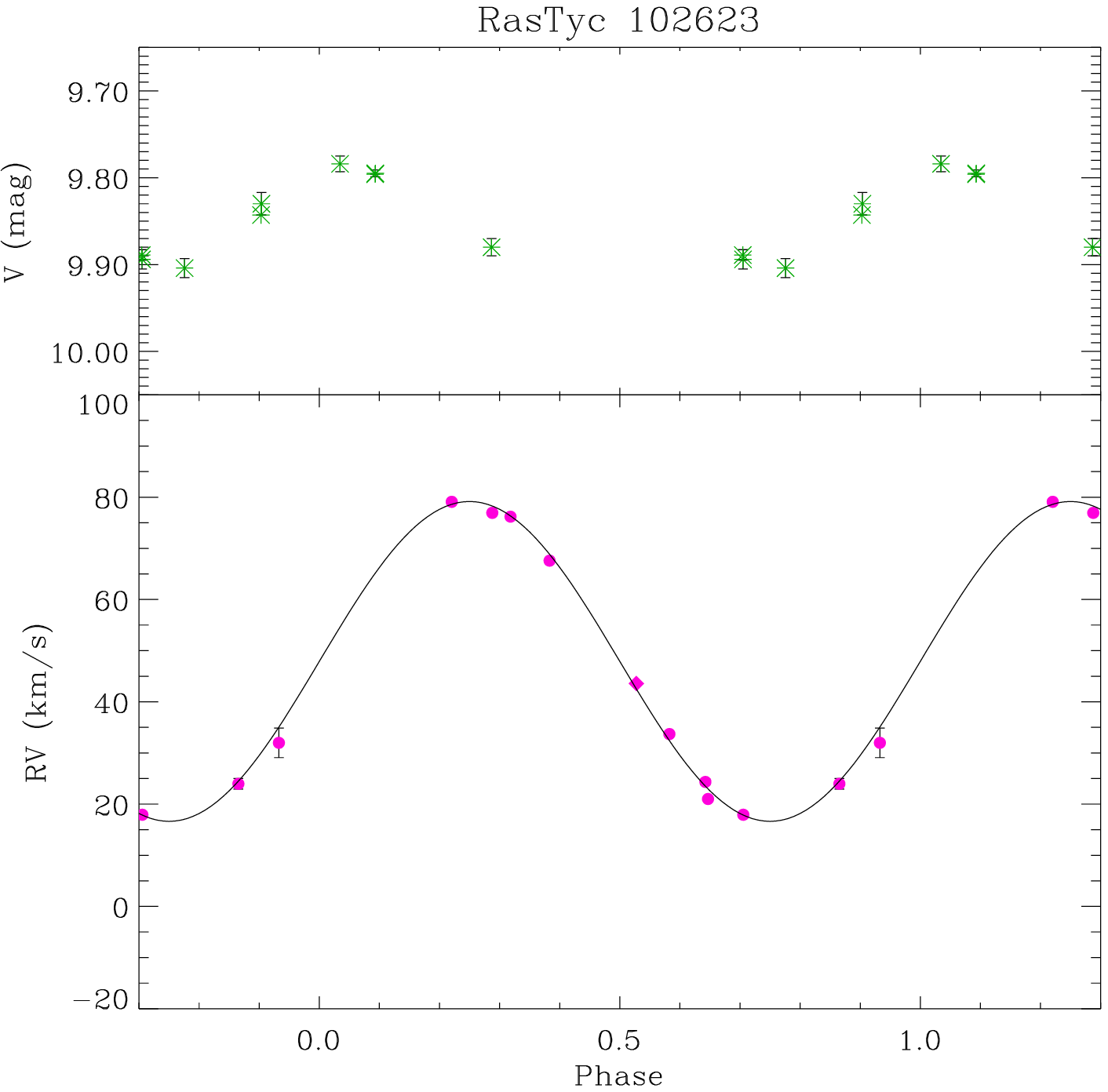}  
\end{center}
\caption{Radial velocity curves (dots= OAC data, diamonds= E{\sc lodie} data) of the new six RasTyc binaries. 
For the three SB2 systems we used filled and open symbols for the primary (more massive) and secondary 
components, respectively. The circular solutions are also represented with solid and dashed lines for the primary 
and secondary components, respectively. The contemporaneous $V$  photometry is displayed, as a function of 
the orbital phase, on the top panel of each box.}
\label{fig:RV}
\end{figure*}

\begin{table}	
\label{Tab-RV}
\caption[Orbital parameters of the new binaries.]
{Orbital parameters of the new binaries.}
\begin{tabular}{llcccc}
\sphline
\it RasTyc  &  \it  Name  &  $P_{\rm orb}$  & 	$\gamma$  &  $k$ (P/S)  &  $M\sin^3i$ (P/S)   \\
        &       &  (days)     &  (km\,s$^{-1}$)  &  (km\,s$^{-1}$)   &  (M$_{\odot}$) \\
\sphline
193137  &  HD~183957	 & 7.954  &   $-$29.0  &   57.5/63.1   & 0.758/0.691  \\
215940  &  OT~Peg        & 1.748  &   $-$27.0  &   16.6/23.2   & 0.007/0.005  \\
221428  &  BD+33\,4462   & 10.12  &   $-$20.9  &   59.2/60.4   & 0.905/0.887  \\
040542  &  DF~Cam        & 12.60  &   $-$19.5  &   22.8        & SB1	      \\
072133  &  V340~Gem	 & 36.20  &   +37.0    &   42.1        & SB1	      \\
102623  &  BD+38\,2140   & 15.47  &   +47.4    &   31.3        & SB1	      \\
\sphline
\end{tabular}
\end{table}

The radial velocity (RV) measurements for the E{\sc lodie} data have been performed onto
the cross-correlation functions (CCFs) produced on-line during the data acquisition.

Radial velocities for OAC spectra were obtained by cross-correlation of
each echelle spectral order of the RasTyc spectra  with that of bright RV 
standard stars. For this purpose the IRAF task {\sc fxcor}, that computes RVs by means of
the cross-correlation technique, was used.

The wavelength ranges chosen for the cross-correlation excluded the H$\alpha$ and 
Na\,{\sc I} D$_2$ lines, which are contaminated
by chromospheric emission and have very broad wings. The spectral
regions heavily affected by telluric lines (e.g. the  O$_2$ lines in the 
$\lambda~6276-\lambda~6315$ region) were also excluded.

The observed RV curves  are displayed in Fig.~\ref{fig:RV}, where, for SB2 systems, 
we used dots for the primary (more massive) components and open circles for the secondary 
(less massive) ones. We initially searched for eccentric orbits  and found in any case very 
low eccentricity values (e.g. $e=0.010$ for HD~183957, $e=0.030$ for RasTyc~221428). Thus, 
following the precepts of \cite{Lucy71}, we adopted $e=0$. The circular solutions are also 
represented in Fig.~\ref{fig:RV} with solid and dashed lines for the primary 
and secondary components, respectively.

The orbital parameters of the systems, orbital period ($P_{\rm orb}$), barycentric velocity ($\gamma$), 
RV semi-amplitudes ($k$) and masses ($M\sin^3i$), are listed in Table~\ref{Tab-RV}, 
where P and S refer to the primary and secondary components of the SB2 systems, respectively.

With the only exception of HD~183957, for which any modulation is visible neither in OAC data nor in TYCHO 
$V_{\rm T}$ magnitudes, all sources show a  photometric modulation well correlated with the orbital period,
indicating a high degree of synchronization. The low amplitude of RasTyc\,215940's  light 
curve and the very low values of $M\sin^3i$ imply a very low inclination of orbital/rotational axis.

\begin{table}
\label{Tab-phys}
\caption[Physical parameters of the new binaries.]
{Physical parameters of the new binaries.}
\begin{tabular}{llcccc}
\sphline
\it RasTyc  &  \it  Name  &  $v\sin i$ (P/S) &  {\it Sp. Type} (P/S) &  $B-V$  &  W$_{\rm LiI}$ (P/S) \\
            &             &  (km\,s$^{-1}$)  &                &         &   (m\AA)   \\
\sphline
193137  &  HD~183957     &   4.0/4.4	 &     K0V/K1V      & 0.84  &	 $< 10$ \\
215940  &  OT~Peg        &   9.2/9.4	 &     G8V/K3-4V    & 0.79  &	 60/--	\\
221428  &  BD+33\,4462   &   9.7/26.5    &     F8V/K0III-IV & 0.70  &	 --/15:	\\
040542  &  DF~Cam        &   35 	 &     K2III	    & 1.14  &	 ---	\\
072133  &  V340~Gem      &   40 	 &     G8III	    & 0.83  &	 70	\\
102623  &  BD+38\,2140   &   11.5	 &     K1IV	    & 1.03  &	 40	\\
\sphline
\end{tabular}
\end{table}

\subsection{Spectral type and $v\sin i$ determination}
\label{sec:Spty}

\begin{figure}[ht]
  \begin{center}
    \includegraphics[width=5.9cm,height=4.3cm]{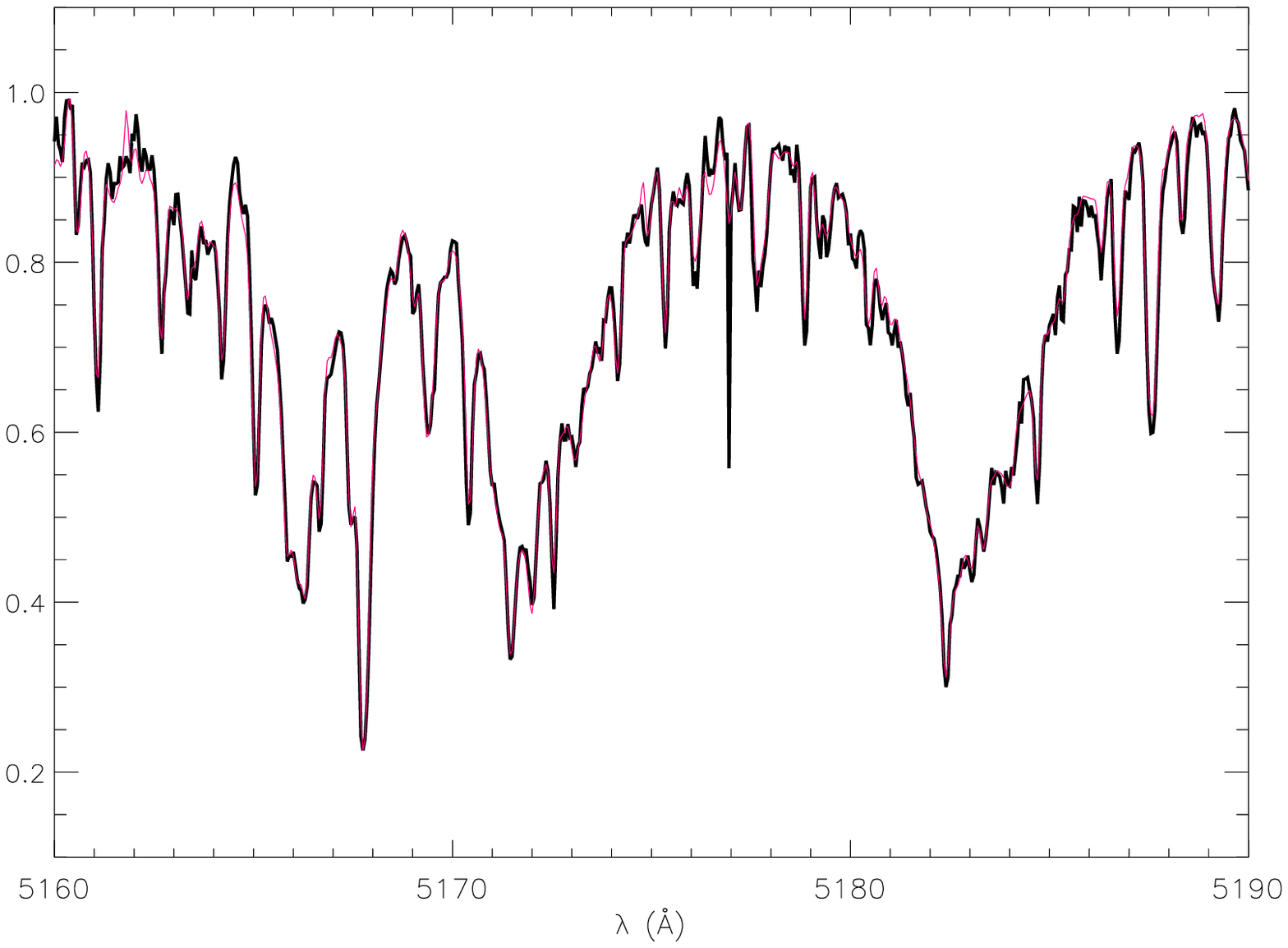} 
    \includegraphics[width=5.9cm,height=4.3cm]{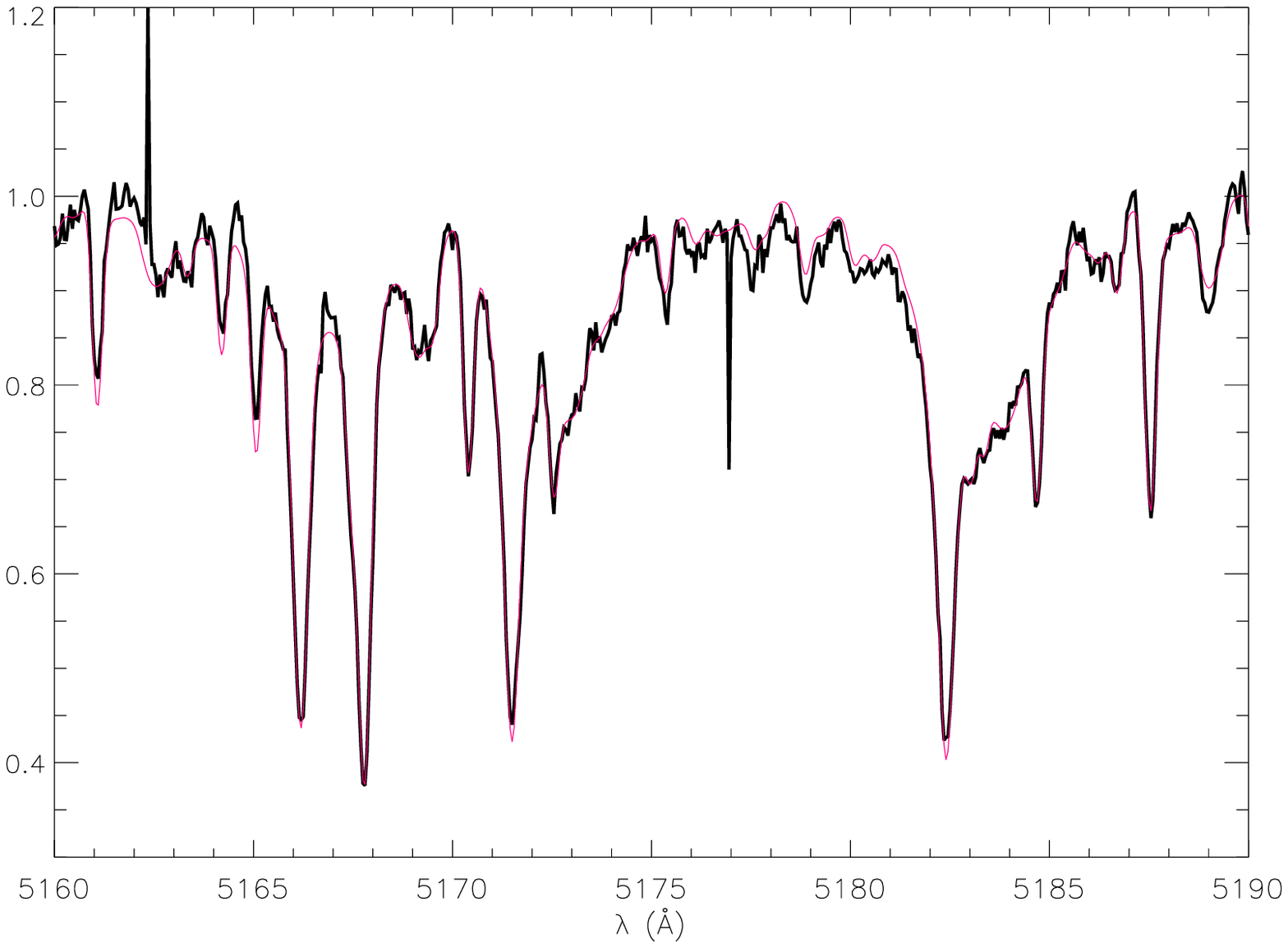} 
  \includegraphics[width=5.9cm,height=4.3cm]{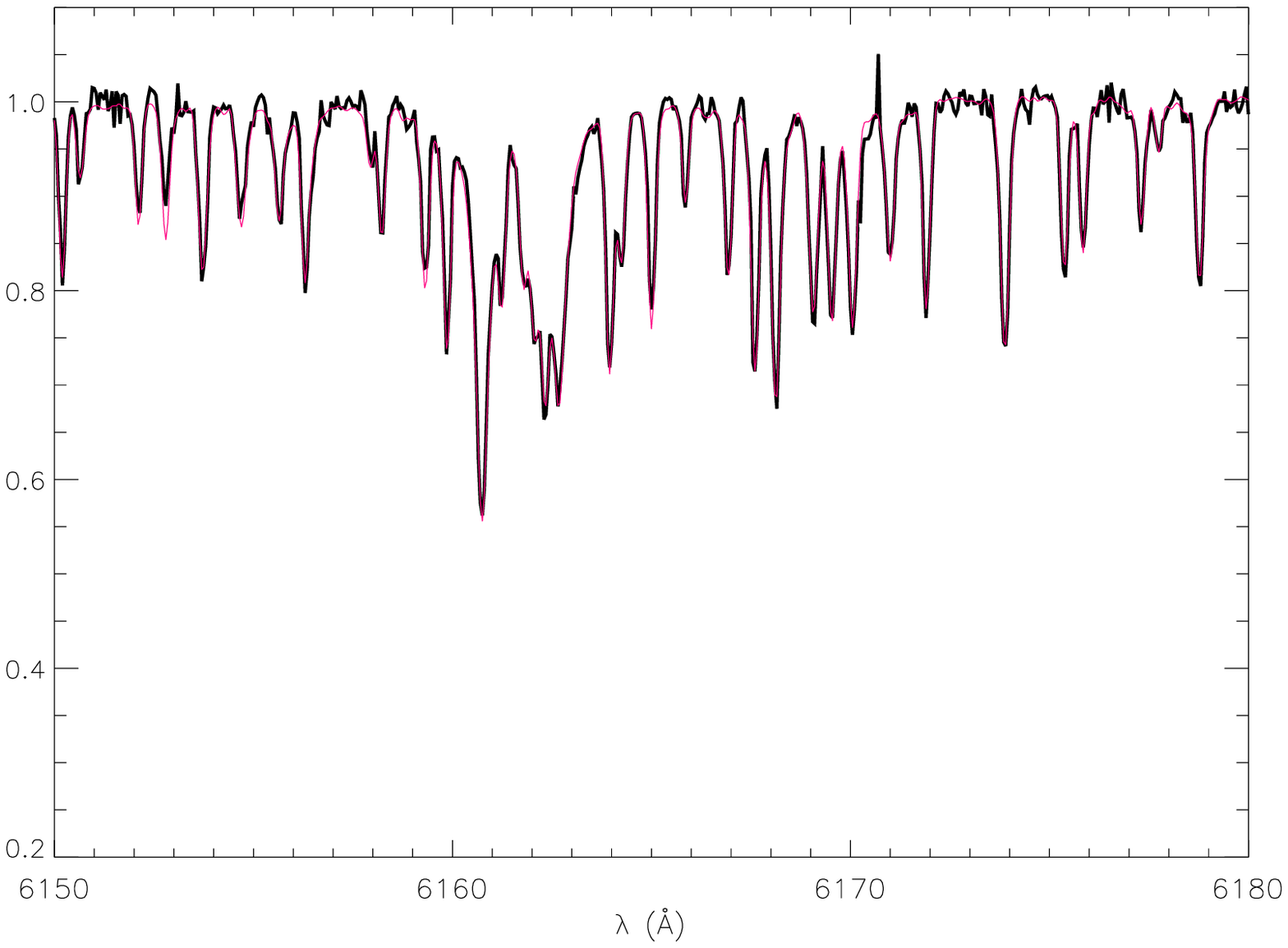}   
  \includegraphics[width=5.9cm,height=4.3cm]{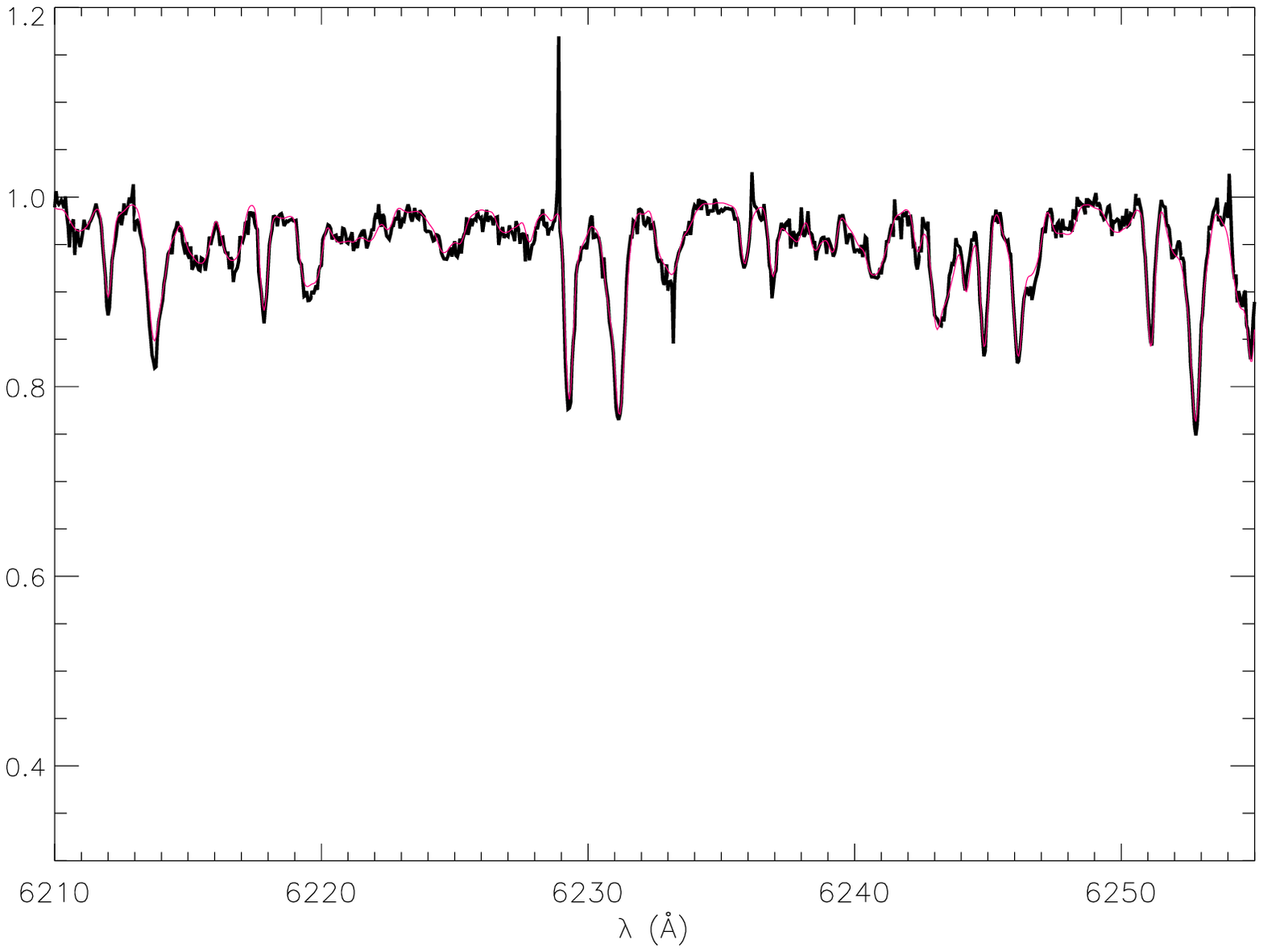}   
  \end{center}
\caption{{\bf Left panels)} Observed E{\sc lodie} spectrum of the SB2 binary HD~183957 (thick black lines) in the 
Mg{\sc i}b (top)and the $\lambda$\,6170 \AA ~(bottom) spectral regions together with the ``synthetic'' spectrum over-plotted 
with a red line. The latter was built up as the weighted sum of two standard star spectra (K0\,V for the primary 
and K1\,V for the secondary one) broadened at the $v\sin i$ of the respective component and Doppler-shifted according 
to the orbital solution and the phase of the observation ($\phi = 0.36$). 
The fractional flux contribution of the primary component is 0.62 and 0.58 at 5170 and 6170 \AA, respectively.  
{\bf Right panels)}  Observed E{\sc lodie} spectrum of RasTyc~221428 (thick black lines) in the Mg{\sc i}b (top)
and in the $\lambda$\,6230 \AA ~(bottom) regions. The ``synthetic'' spectrum (F8\,V + K0\,III-IV) is 
displayed with red lines in the same boxes. The excellent agreement of the observed and synthetic spectra is apparent.}
\label{fig:synth}
\end{figure}

The projected rotational velocity, $v\sin i$, was measured using the E{\sc lodie} CCFs and the calibration 
relation between CCF width and $v\sin i$ proposed by \cite{Queloz98} and is reported in Table 2. 
The lowest rotation rate ($v\sin i \simeq$\,4 km\,s$^{-1}$) has been detected for both components 
of HD~183957, which display also the lowest H$\alpha$ activity among the six sources.

For SB1 systems observed with E{\sc lodie} we have determined effective temperatures and gravity (i.e. spectral 
classification) by means of the TGMET code, available at OHP (\cite{Katz98}). We have also used ROTFIT, a code 
written by one of us (\cite{Frasca03}) in IDL (Interactive Data Language, RSI), which simultaneously find 
the spectral type and the $v\sin i$ of the target by searching, into a library of standard star spectra, for 
the standard spectrum which gives the best match (minimum of the residuals in the difference) of the target one, 
after the rotational broadening. As a standard star library, we used a sub-sample of the entire TGMET list, 
composed of 87 stars well distributed in effective temperature and gravity and in a suitable range of metallicities. 
The standard star spectra were retrieved from the E{\sc lodie} Archive (\cite{Prugniel01}). 
The ROTFIT code was also applied to the OAC spectra, using standard star spectra acquired at OAC with the same 
instrumental setup. This was especially advantageous for DF~Cam, for which we have no E{\sc lodie} spectrum.  
We have performed an accurate spectral classification also for the components of SB2 systems, thanks to
another IDL code made by us, COMPO2, which searches for the best combination of standard-star spectra
able to reproduce the observed spectrum of the SB2 system. We give the radial velocity and $v\sin i$ of the 
system components as input parameters and the code finds, for the selected  spectral region, the spectral
types and fractional flux contributions which better reproduce the observed spectrum, i.e. which minimize the 
residuals in the collection of difference, $observed - synthetic$ spectra.  
Results for different E{\sc lodie} orders from blue to red wavelengths are in very good agreement 
(Fig.~\ref{fig:synth}).

We found only two binaries composed of main sequence stars, while the remaining systems contain an evolved
star (giant or sub-giant) and can be classified as new RS~CVn systems (see Table 2).  

\subsection{H$\alpha$ emission and Lithium content}
\label{sec:Halpha}

\begin{figure}[ht]
  \begin{center}
    \includegraphics[width=5.7cm,height=4cm]{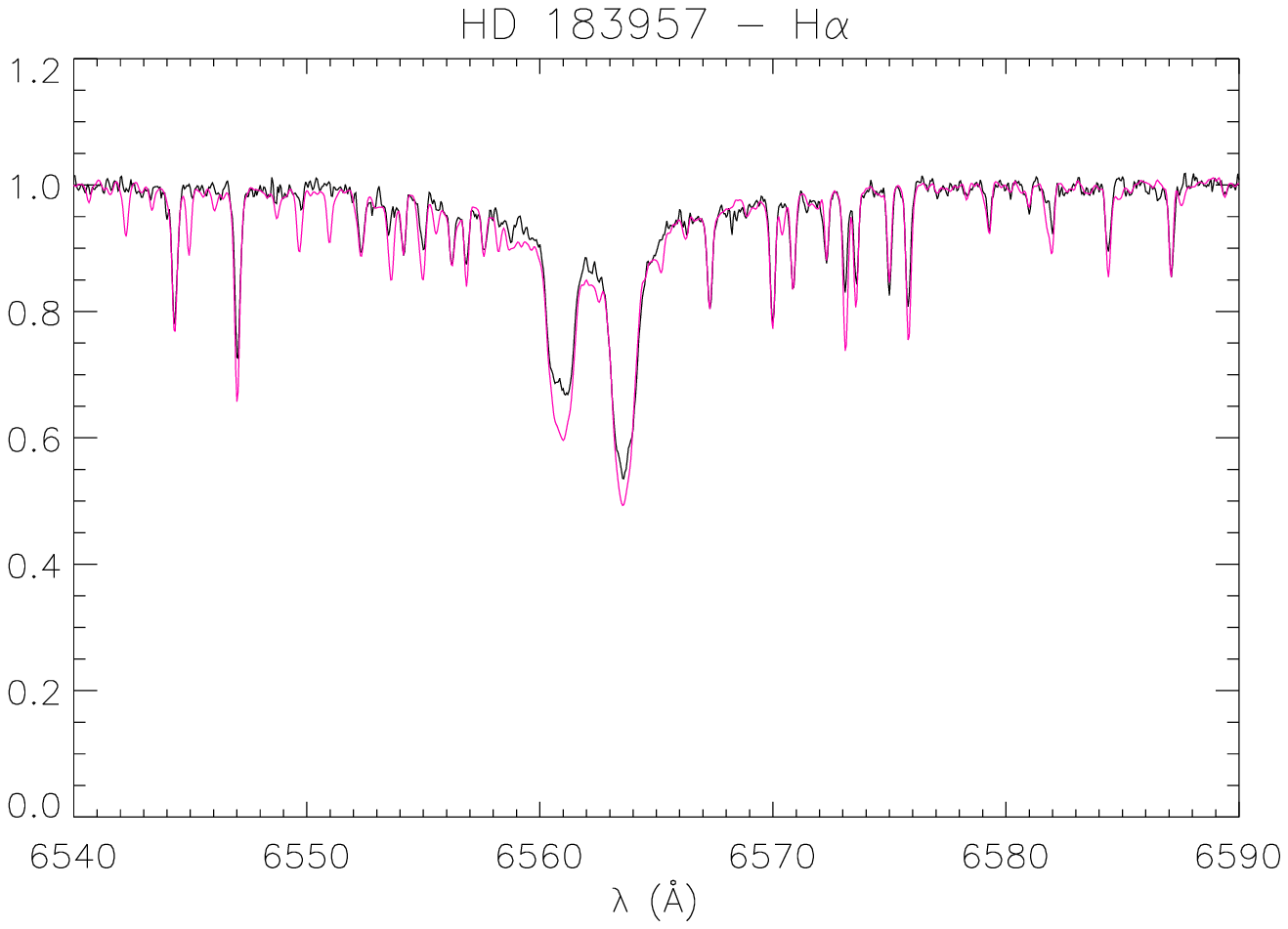}
  \includegraphics[width=5.7cm,height=4cm]{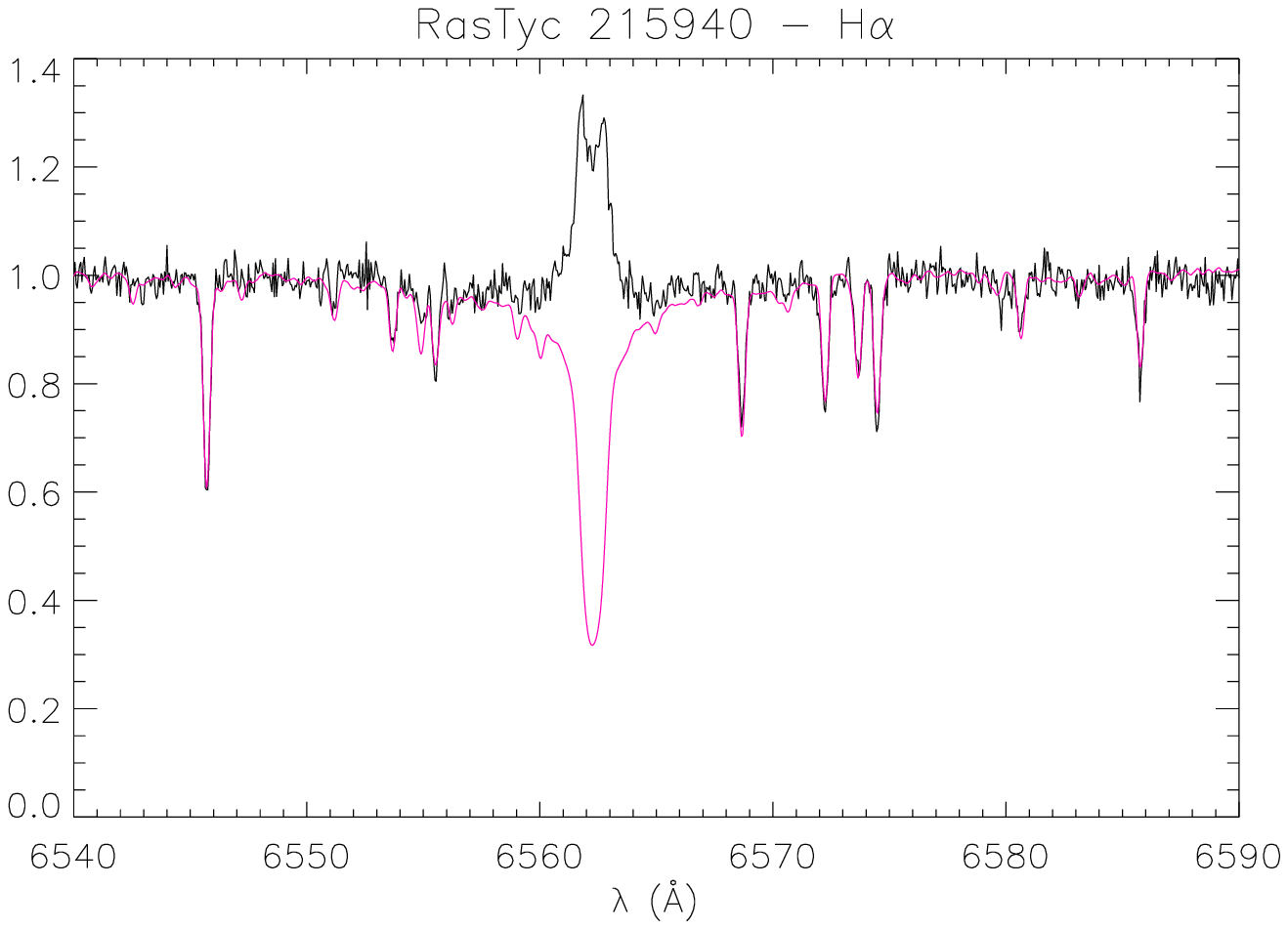}	
    \includegraphics[width=5.7cm,height=4cm]{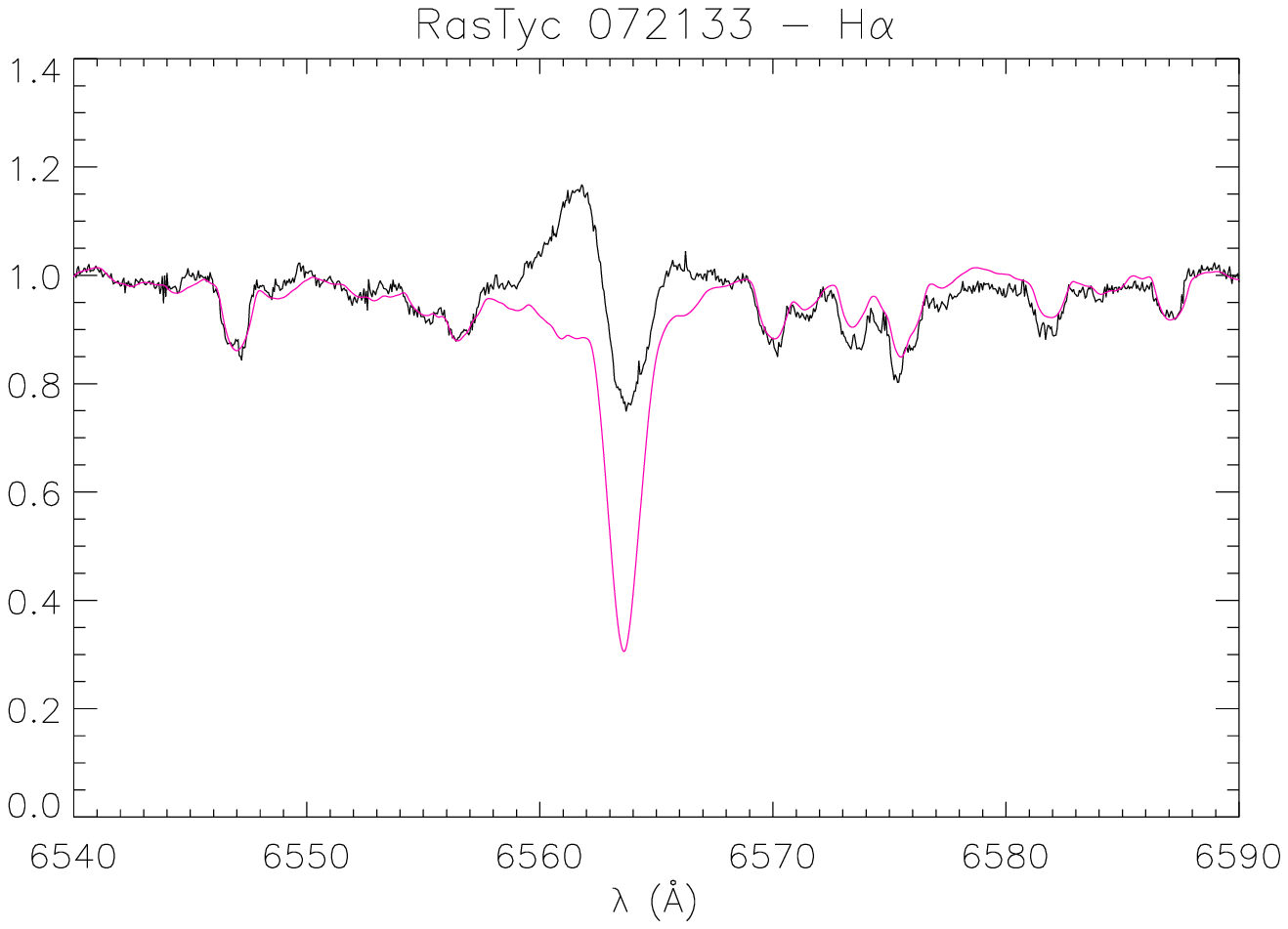}	
  \includegraphics[width=5.7cm,height=4cm]{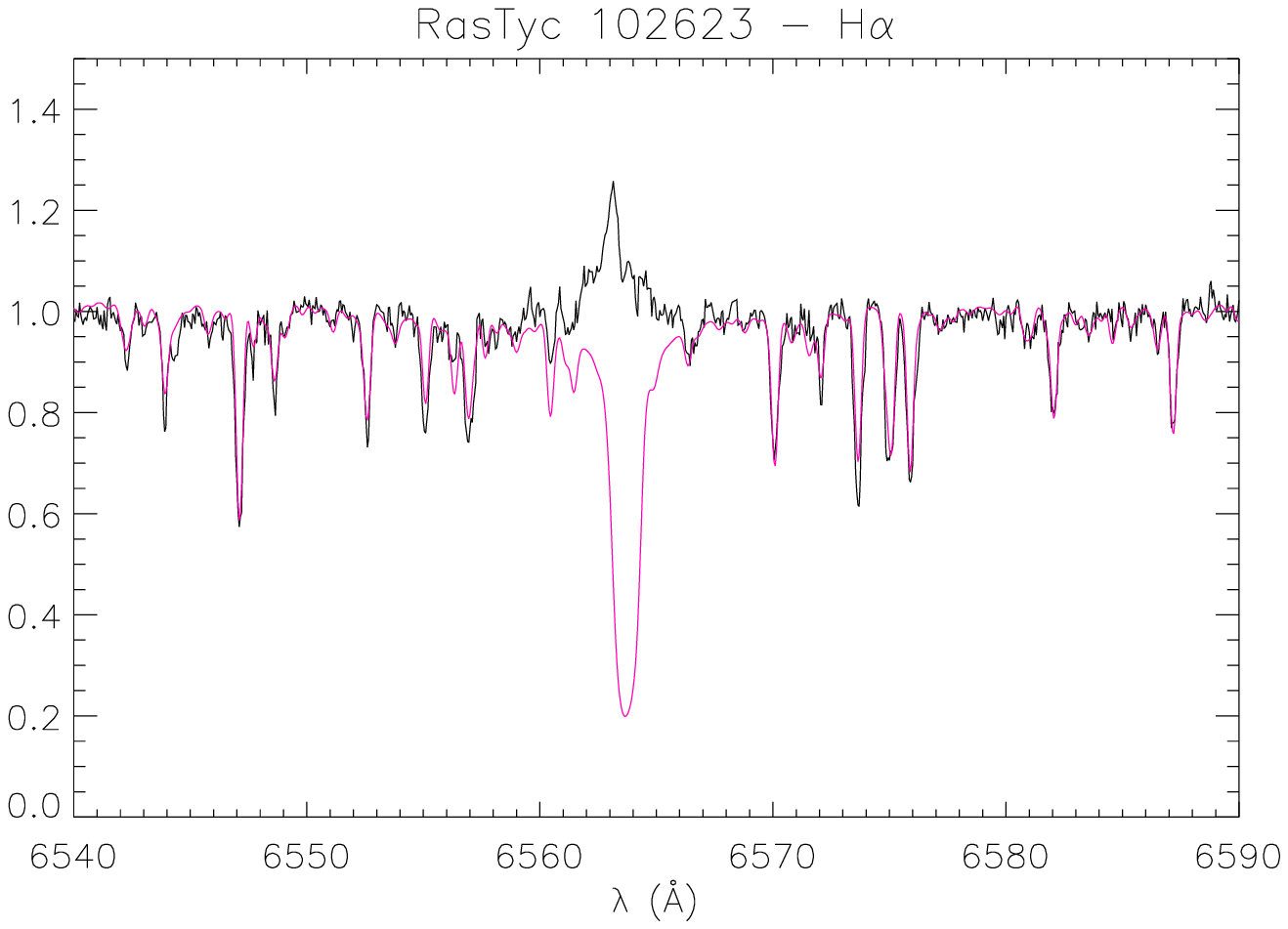}	
  \end{center}
\caption{E{\sc lodie} spectra of HD~183957, RasTyc 215940, 072133, and 102623 in the H$\alpha$ 
region (black lines). The inactive 
templates built up with rotationally broadened E{\sc lodie} Archive spectra are displayed with red lines. 
The two components of HD~183957 show only a small filling of their H$\alpha$ profiles, while RasTyc 072133
displays a double-peaked H$\alpha$ emission profile. RasTyc 215940 and 102623 display instead a pure emission 
H$\alpha$ profile.}
\label{fig:Halpha1}
\end{figure}

\begin{figure}[ht]
  \begin{center}
    \includegraphics[width=5.7cm,height=9cm]{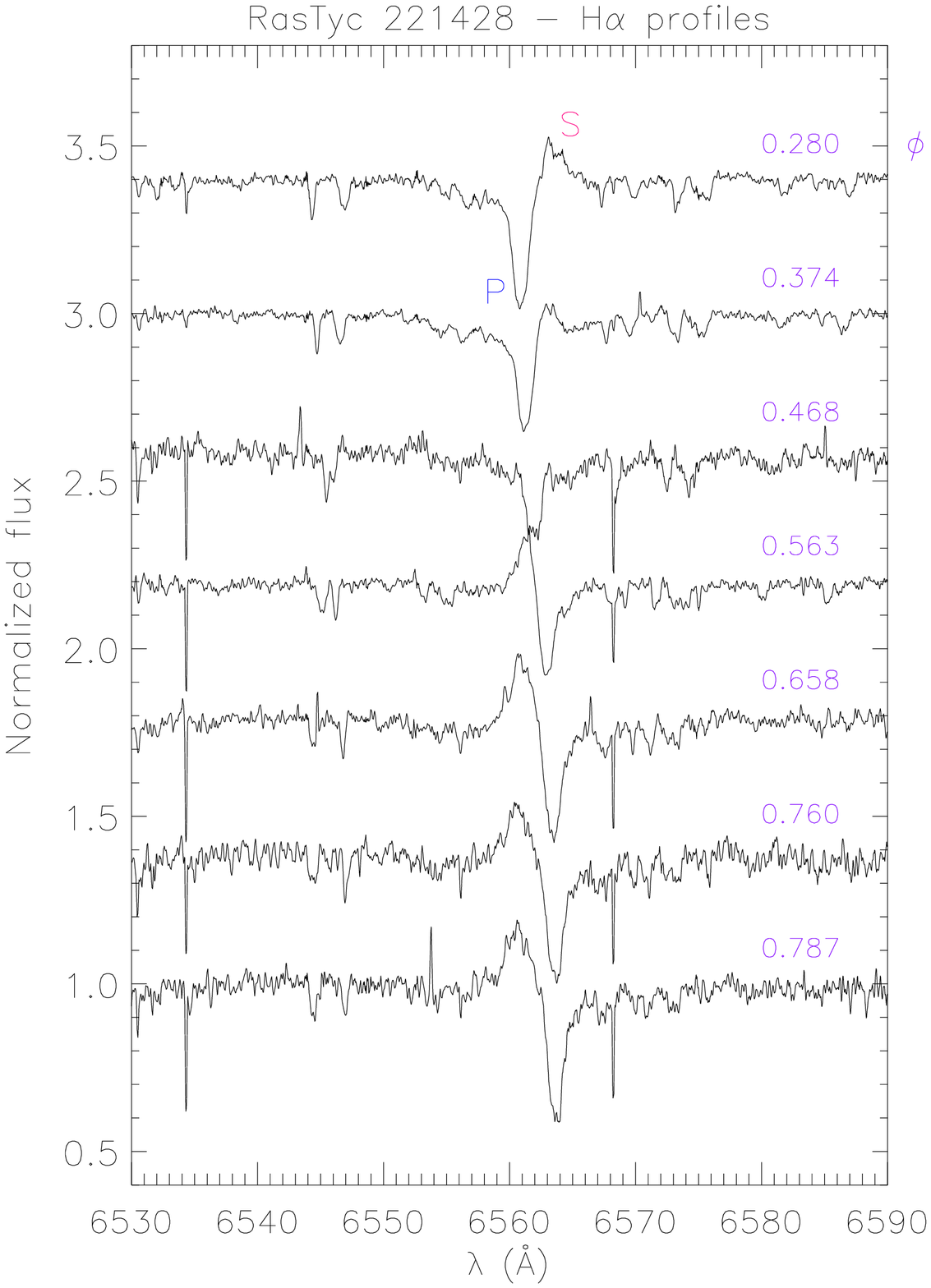}  
  \includegraphics[width=5.7cm,height=9cm]{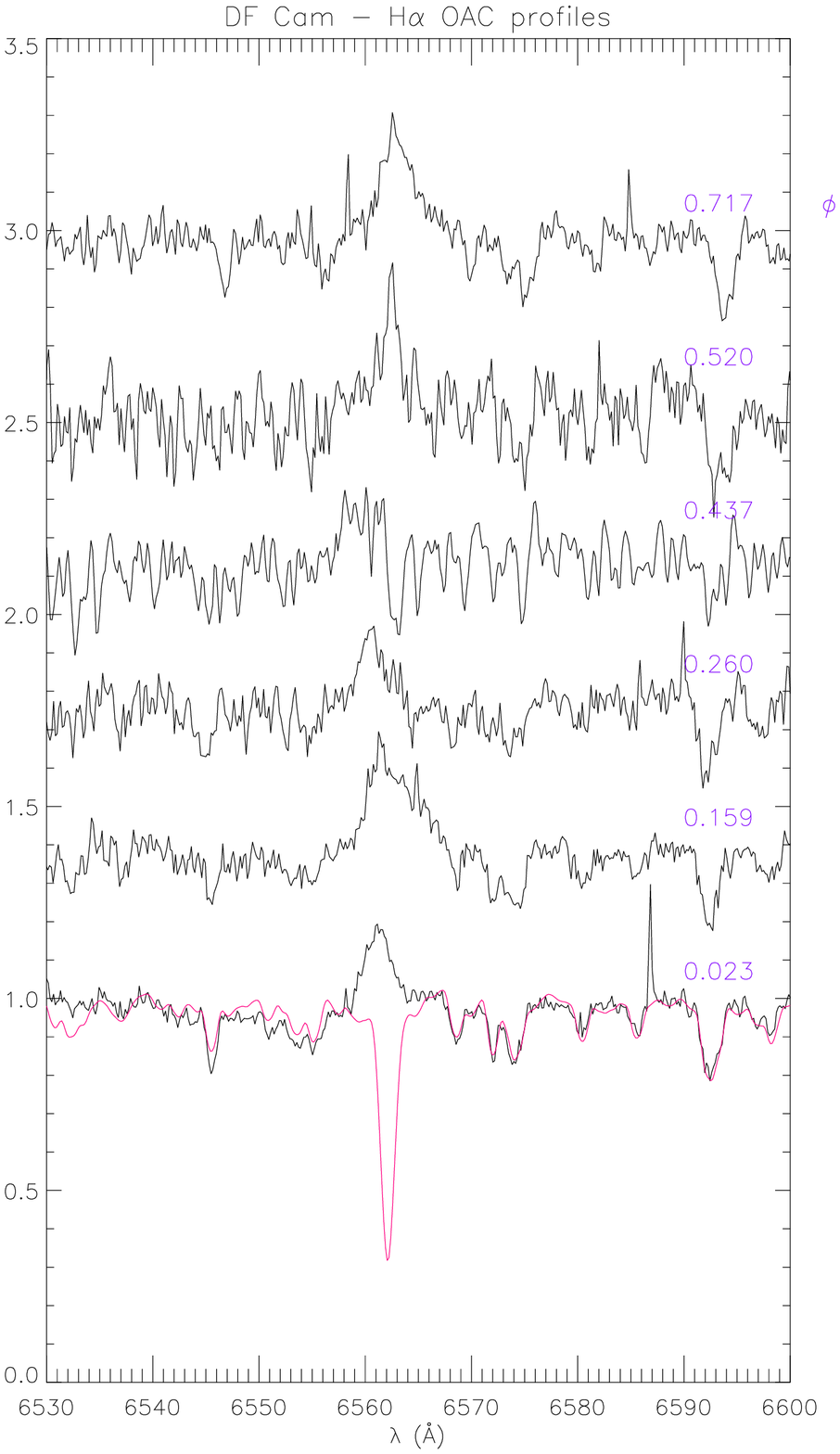}
  \end{center}
\caption{{\bf Left panel)} Sample of E{\sc lodie} spectra of RasTyc~221428 in the H$\alpha$ region at different
orbital phases. The cooler K0~III-IV component shows a H$\alpha$ line always in emission with 
a stronger intensity around phase 0.7, while the hotter F8~V component displays a normal H$\alpha$
absorption profile.
{\bf Right panel)} OAC spectra of DF~Cam in the H$\alpha$ region at different orbital phases.
The H$\alpha$ intensity and shape varies with the orbital/rotational phase.}
\label{fig:Halpha2}
\end{figure}

The H$\alpha$ line is one of the most important and easily accessible indicators of chromospheric activity.
Only the very active stars show the H$\alpha$ line as an emission feature always above the local continuum, 
while in the spectra of less active objects only a filled-in absorption line is observed. 
The detection of the chromospheric emission contribution filling in the line core is hampered by 
the complexity of the observed spectra of double-lined binaries in which the spectral lines of both 
components, shifted at different wavelengths according to the orbital phase, are simultaneously seen.
Therefore a comparison with an ``inactive" template, composed of two stellar
spectra mimicking the two system components in absence of activity, is needed to emphasize
the H$\alpha$ chromospheric emission. 

The inactive templates have been built up with rotationally broadened E{\sc lo\-die} archive spectra 
(HD~165341, K0\,V and HD~10476, K1\,V for the primary and secondary component of HD~183957, respectively; 
$\gamma$~Cep, K1\,IV for 102623; $\delta$~Boo, G8\,III for 072133; HD~17382, K1\,V for 215940) or with OAC spectra
of $\alpha$~Ari (K2\,III), for DF~Cam, acquired during the observing campaigns.

The two components of HD~183957 show only a small filling of their H$\alpha$ profiles (Fig.~\ref{fig:Halpha1}),
i.e. a moderate activity level. This is in agreement with the low $v\sin i$ values indicating a system 
composed of two early-K stars which are probably synchronously rotating with a period of about 8 days. 
The filling in of the H$\alpha$ line of both components of HD~183957 is lower than that found for HD~166 (\cite{Biazzo04}),
a K0\,V star belonging to the Local Association (100-300 Myrs) with a rotation period of 6.2 days.
  
The other RasTyc stars display H$\alpha$ emission profiles with a variety of shapes, going from a simple 
symmetric emission profile (102623) to a strong emission line with central reversal (215940). It has been also observed
a very broad, complex feature with a filled-in core and an emission blue wing (072133). 
A H$\alpha$ profile similar to that displayed by the latter star has been sometimes 
observed in some long-period RS CVn's, like HK Lac (e.g. Catalano \& Frasca \cite{Cata94}). RasTyc 072133 was 
classified as a semi-regular variable after Hipparcos, but it displays all the characteristics of a 
RS~CVn SB1 binary.

The E{\sc lodie} spectra of RasTyc 221428 in the H$\alpha$ region show that the cooler K0~III-IV 
component displays a H$\alpha$ line always in emission with a stronger intensity around phase 0.7
(Fig.~\ref{fig:Halpha2}). The OAC spectra of DF~Cam always display a pure H$\alpha$ emission line, 
whose intensity  varies with the orbital/rotational phase (Fig.~\ref{fig:Halpha2}). Similarly to RasTyc 
072133, DF~Cam, considered as a semi-regular variable after Hipparcos photometry, is very likely an 
active binary of the RS~CVn class.

The equivalent width of the lithium $\lambda$6708 line, $EW_{\rm Li}$, was measured on the 
E{\sc lodie} spectra.  For the three sources for which we were able to detect and measure 
$EW_{\rm Li}$ (Table~2), we have deduced a lithium abundance, $\log N(Li)$, in the range 1.3--1.8, 
according to \cite{Pav96} NLTE calculations. Such values of lithium abundance indicate ages of at 
least a few hundreds Myrs, so that
the X-ray emission and the fairly strong H$\alpha$ chromospheric activity detected in these systems 
should be essentially the effect of the rather fast stellar rotation due to the spin/orbit synchronization.

\begin{acknowledgments}
We are grateful to the members of the staff of OHP and OAC observatories for their support and help
with the observations.
This research has made use of SIMBAD and VIZIER databases, operated at CDS, 
Strasbourg, France.
\end{acknowledgments}

\end{document}